\documentclass[letterpaper]{article}
\usepackage{acl2013}
\usepackage{times}
\usepackage{helvet}
\usepackage{courier}
\usepackage{amsfonts}
\usepackage{url}
\usepackage{booktabs}
\usepackage{latexsym}
\usepackage{amsmath}
\usepackage{amssymb}
\usepackage{multirow}
\usepackage{color}
\usepackage{graphicx}
\usepackage{verbatim}

\newcommand{\todo}[1]{\textcolor{red}{#1}}

\DeclareMathOperator*{\argmax}{arg\,max}

\title{I Wish I Didn't Say That!  Analyzing and Predicting Deleted Messages in
Twitter}
\author{Sa\v sa Petrovi\' c\\
School of Informatics\\
University of Edinburgh\\
{\tt sasa.petrovic@ed.ac.uk}
\And
Miles Osborne\\
School of Informatics\\
University of Edinburgh\\
{\tt miles@inf.ed.ac.uk}
\And
Victor Lavrenko\\
School of Informatics\\
University of Edinburgh\\
{\tt vlavrenk@inf.ed.ac.uk}
}

\date{}

\begin{document}
\maketitle

\begin{abstract}

Twitter has become a major source of data for social media researchers.  One
important aspect of Twitter not previously considered are {\em deletions} --
removal of tweets from the stream.  Deletions can be due to a multitude of
reasons such as privacy concerns, rashness or attempts to undo public
statements.  We show how deletions can be  automatically predicted
ahead of time and  analyse which tweets are likely to be deleted and how.

%s that  users can delete tweets at any time in the future and researchers need to
%honor these deletion requests.  We show that published research on Twitter
%which respects these enforced deletion requests can be non-reproducible.
%Additionally we observe that users retract tweets .  All of these
%concerns can be tackled from the novel perspective of predicting when a tweet
%is likely to be deleted in the future.  We report on the first results for this
%task, explaining some of its characteristics and grounding it in human
%performance.  

\end{abstract}
%	One of the biggest problems with conducting research on Twitter is the fact
%	that it is hard to replicate previous results due to some of the data no
%	longer being available.  This unavailability is a direct consequence of
%	tweets being deleted from the stream, and it means that two researchers can
%	obtain different results using the same method on the same data, but
%	crawled at different times.  On the other hand, when Twitter users delete
%	their tweets they are probably unaware of the fact that thousands of
%	applications automatically crawl their tweets, and that not posting the
%	tweet in the first place is the only guaranteed method that no one will
%	ever read it.  We introduce a new task for Twitter: predicting which
%	messages will be deleted from the stream.  This task addresses all of the%
%	above problems (and more), and thus has potential to substantially improve
%	the quality of research on Twitter, as well as user's experience.  We
%	perform the first analysis of deleted messages on Twitter and find that,
%	for example, messages written between 1am and 3am are much more likely to
%	be deleted.  We use a machine learning approach for predicting deletions
%	and show that it significantly outperforms humans on this task.
	
%\end{abstract}

\section{Introduction}

In recent years, research on Twitter has attracted a lot of interest,
primarily due to its open API that enables easy collection of data. 
The belief that tweets contain useful information has
lead to them being used to predict many real-world quantities.  For example,
tweets have been used to predict
elections~\cite{tumasjan10predicting,oconnor10tweets}, stock market
movement~\cite{bollen11twitter}, and even flu outbreaks~\cite{ritterman09using}.
Twitter forbids distribution of raw tweets and their terms of service
insist that any tweet collection must honor post-hoc {\em deletion}
requests. That is, at any point in the future a user can issue a
request to Twitter to delete a tweet.
Predicting when a tweet is likely to be retracted by a user has
important applications:
\begin{itemize}
\item  {\em Security.} Twitter has become so ubiquitous that users often do not
	consider the potential confidentiality implications before they tweet.
\item  {\em Regret.} Users might post an inappropriate or offensive tweet in
	the heat of the moment, only to regret it later.
\item {\em Public scrutiny.} High profile politicians at times tweet content that
	they later withdraw.
\end{itemize}
%MO:  this should perhaps be moved later
%
%For example, the TREC microblog
%track\footnote{\url{http://trec.nist.gov/data/tweets/}} distributes a set of
%queries and corresponding relevance judgments, but any of the relevant
%documents (or even all of them) might be deleted at any point in the
%future.
Here we report on the first results of automatically predicting if tweets will be deleted
  in the future. We also analyse why tweets are deleted.

\section{Related Work}

Predicting deleted messages has been previously addressed in the context of
emails~\cite{dabbish03marked,dabbish05understanding}.  For example,
\cite{dabbish03marked} found that the most important factors affecting the
chances of an email being deleted are the past communication between the two
parties and the number of recipients of the email.  However, it should be clear
that people use tweets in very different ways to using email.  
The most similar work to ours is the recent analysis of censorship in Chinese
social media~\cite{bamman12censorship}.  The problem examined there is that of
the government deleting posts in the Chinese social media site Sina Weibo
(Chinese equivalent of Twitter).  The authors analyze different terms that are
indicative of a tweet being deleted and the difference between appearance of
certain political terms on Twitter and on Sina Weibo.  However, they make no
attempt to predict what will be deleted and only briefly touch upon deleted
messages in Twitter.  While the main reason for deletion in Sina Weibo seems to
be government censorship,\footnote{These results were also confirmed in
\cite{tschang12analysis}.} there is no known censorship on Twitter, and
thus the reasons for deletion will be quite different.  To the best of our
knowledge, we present the first analysis of deleted messages on Twitter.

\begin{comment}
	child12blogging
	---------------
	- they conduct a survey where they ask 318 bloggers their reasons for
	deleting a post.
	- the bloggers used different platforms, but none of them seemed to use
	Twitter.  the only
	condition was that they are active (post at least once a week)
	- authors find six types of bloggers: self-centric, utilitarian, planning,
	sharing, protective, and unworried bloggers
	- they also find seven reasons why bloggers delete posts: conflict
	management, protection of personal identity, fear of retribution,
	employment security, impression management, emotional regulation, and
	relational clensing
	- how are we different: we try to *predict* if a tweet will be deleted and
	quantitatively measure deletion practices on Twitter.  More importantly,
	none of the bloggers used Twitter.
\end{comment}

\section{Task Description}

There are several ways in which a tweet can be deleted.  The most obvious way
is when its author explicitly deletes it (this is usually done by clicking on a
\emph{Delete} button available in most Twitter clients).  Another way that a
tweet becomes effectively deleted is when a user decides to make his tweets
protected.  Although the user's tweets are still available to read for his
friends, no one else has access to them any more (unless the user decides to
make them public again).  Finally, the user's whole account might be deleted
(either by their own choice or by Twitter), meaning that all of his tweets are
also deleted.  In the public streaming API, Twitter does not differentiate
between these different scenarios, so we collapse them all into a single task:
for each tweet predict if it will be deleted, by either of the aforementioned
ways.  

\subsection{Example Deleted Tweets}

Table~\ref{tab:deleted-example} shows some examples of the various types of
deleted tweets that we have discussed (identifiable information has been
replaced by ***).  Although we can never be sure of the true reason behind
someone deleting a tweet, a lot of the time the reason is fairly obvious.  For
example, it is very likely that tweet 1 was deleted because the author
regretted posting it due to its somewhat inappropriate content.  On the other
hand, tweet 2 was most likely posted by a spammer and got deleted when the
author's account was deleted.  Tweet 3 is probably an example of deleting a
tweet out of privacy concerns -- the author posted his email publicly which
makes him an easy target for spammers.  The fourth tweet is an example of a
deleted tweet authored by a Canadian politician (obtained from the website
\url{politwitter.ca/page/deleted}).  Finally, tweet 5 is an example of a false
rumour on Twitter.  This tweet was retweeted many times right after it was
posted, but once it became clear that the news was not true, many users deleted
their retweets.

\begin{table*}
	\centering
	\begin{tabular}{l l}
		\toprule
	1 & \footnotesize{Another weekend without seeing my daughters-now if I'd shot my ex when we
	split I would of been out by now,}\\
	  & \footnotesize{missed opportunity :(} \\
	2 & \footnotesize{Get more followers my best friends? I will follow you back if you
	follow me - http://***} \\
	3 & \footnotesize{@*** yeah man email the contract to ***@gmail.com \dots This has been
	dragged out too long big homie}\\
	4 & \footnotesize{Gov must enforce the Air Canada Act and save over 2,500 jobs. @***
	http://*** \#ndpldr} \\
	5 & \footnotesize{BREAKING: URGENT: News spreading like wildfire, BASHAR AL-ASSAD HAS
	ESCAPED \#SYRIA!} \\
	  & \footnotesize{We're waiting for a confirmation} \\
	\bottomrule
	\end{tabular}
	\caption{Examples of tweets that have been deleted.}
	\label{tab:deleted-example}
\end{table*}

\section{Predicting when Tweets will be Deleted}
\label{sec:experiments}
We now show the extent to which tweet deletion can be automatically
predicted.

\subsection{Data}

We use tweets collected from Twitter's streaming API during January 2012.  This
data consists of 75 million tweets, split into a training set of 68 million
tweets and a test set of about 7.5 million more recent tweets (corresponding
roughly to tweets written during the last three days of January 2012).  A tweet
is given the label 1, meaning it was deleted, if the notice about its deletion
appeared in the streaming API at any time up to 29th February 2012.  Otherwise
we consider that the tweet was not deleted.  In total, 2.4 million tweets in
our dataset were deleted before the end of February.

\subsection{Features}
\label{sec:features}

We use the following features for this task:
\begin{itemize}

	\item \emph{Social} features:  user's number of friends, followers,
		statuses (total number of tweets written by a user), number of lists
		that include the user, is the user verified, is the tweet a retweet, is
		the tweet a reply.  Additionally, we include the number of hashtags,
		mentions, and links in the tweet under social features, even though
		they are not strictly ``social''.  We do this because these features
		are dense, and thus much more similar to other dense features (the
		``real'' social features) than to sparse features like the author and
		text features.

\item \emph{Author} features: user IDs, 
\item \emph{Text} features: all the words in the tweet.
\end{itemize}

Because of the user IDs and lexical features, the feature set we use is
fairly large.  In total, we have over 47 million features, where 18 million
features are user IDs, and the rest are lexical features (social features
account for only about a dozen of features).  We do not use features
like user's time zone or the hour when the tweet was written.  This is because
our preliminary experiments showed that these features did not have any effect
on prediction performance, most likely because the author and text features
that we use already account for these features (e.g., authors in different time
zones will use different words, or tweets written late at night will contain
different words from those written in the morning).

\subsection{Learning Algorithm}

In all our experiments we use a support vector machine
(SVM)~\cite{cortes95support} implemented in Liblinear~\cite{fan08liblinear}.
We note that while SVMs are generally found to be very effective for a wide
range of problems, they are not well suited to large-scale streaming problems.
A potential limitation is the fact that they require batch training, which can
be prohibitive both in terms of space and time when dealing with large
datasets.  Because of this, we also explored the use of the passive-aggressive
(PA) algorithm~\cite{crammer06online}, which is an efficient, online,
max-margin method for training a linear classifier.  Thus, we also
present results for PA as an alternative for cases where the data is simply too
big for an SVM to be trained.

\subsection{Results}

We formulate predicting deletions as a binary classification task -- each
tweet is assigned a label 0 (will not be deleted) or 1 (will be deleted).
Because the two classes are not equally important, i.e., we are normally more
interested in correctly predicting when something will be deleted than
correctly predicting when something will not be deleted, we use the $F_1$ score
to measure performance.  $F_1$ score is standard, e.g., in information
retrieval, where one class (relevant documents) is more important than the
other.

Results are shown in Table~\ref{tab:deletion-results}.  The random baseline
randomly assigns one of the two labels to every tweet, while the majority
baseline always assigns label 1 (will be deleted) to every tweet.  We can see
from the absolute numbers that this is a hard task, with the best $F_1$
score of only 27.0.  This is not very surprising given that there are many
different reasons why a tweet might be deleted.  Additionally, we should keep
in mind that we work on all of the crawled data, which contains tweets in
nearly all major languages, making the problem even harder (we are trying to
predict whether a tweet written in \emph{any} language will be deleted).
Still, we can see that the machine learning approach beats the baselines by a
very large margin (this difference is statistically significant at $p=0.01$).  
Further improving performance in this task will be the focus of future work and
this should enable researchers to distribute more stable Twitter datasets.

We mentioned before that using an SVM might be prohibitive when dealing with
very large datasets.  We therefore compared it to the PA algorithm and found
that PA achieves an $F_1$ score of 22.8, which is 4.2 points lower than the SVM
(this difference is significant at $p=0.01$)  However, the SVM's gain in
performance might be offset by its additional computational cost -- PA took 3
minutes to converge, compared to SVM's 8 hours, and its memory footprint was
two orders of magnitude smaller.  Because efficiency is not our primary concern
here, in the rest of the paper we will only present results obtained using SVM,
but we note that the results for PA showed very similar patterns.  

%We have also tried using second-order features of the type (\emph{user ID}
%$\times$ \emph{lexical features}).  Previously, \cite{daume07frustratingly}
%showed that such features tend to help in similar prediction tasks.  We found
%that these features improved the performance of PA, but not to the extent that
%it outperformed SVM without such features.  Unfortunately, we were not able to
%use the second-order features in SVM because the Liblinear implementation did
%not scale to this huge number of features.  Nevertheless, based on our
%experiments with PA and the previous research that used such features, we would
%expect them to also improve the performance of SVMs.

\begin{table}
	\centering
	\begin{tabular}{l c}
		\toprule
	    & $F_1$ \\
		\midrule
		Random baseline    & 5.8  \\
		Majority baseline  & 6.0  \\
		All features (SVM) & 27.0 \\
		All features (PA)  & 22.8 \\
		%All + second-order features (PA) & 25.3 \\
		\midrule
		Social features    & 3.8  \\
		%Lexical features   & 10.7 \\
		%Lexical features   & 13.1 \\
		Lexical features   & 10.9 \\
		User IDs           & 12.2 \\
		%User IDs           & 6.0  \\
		\bottomrule
	\end{tabular}
	\caption{Results for predicting deleted tweets.}
	\label{tab:deletion-results}
\end{table}

To get more insight into the task, we look at how different feature types
affect performance.  We can see from the last three rows of
Table~\ref{tab:deletion-results} that social features alone achieve very poor
performance.  This is in contrast to other tasks on Twitter, where social
features are usually found to be very helpful (e.g., \cite{petrovic11rt} report
$F_1$ score of 39.6 for retweet prediction using only social features).
Lexical features alone achieved reasonable performance, and the best
performance was achieved using user ID features.  This suggests that some users
delete their tweets very frequently and some users almost never delete their
tweets, and knowing this alone is very helpful.  Overall, it is clear that
there is benefit in using all three types of features, as the final performance
is much higher than performance using any single feature group.

We performed ablation experiments where we removed social features from the
full set of features one at a time and measured the change in performance.  We
found that the only two features that had an impact greater than 0.1 in $F_1$
were the number of tweets that the user has posted so far (removing this
feature decreased $F_1$ by 0.2), and is the tweet a retweet (removing
this feature decreased $F_1$ by 0.16).  This is interesting because the number
of statuses is usually not found to be helpful for other prediction tasks on
Twitter, while the followers number is usually a very strong feature, and
removing it here only decreased $F_1$ by 0.07.

\begin{comment}
\begin{table}
	\centering
	\begin{tabular}{l r}
		\toprule
		Feature & Effect on $F_1$ \\
		\midrule
		%Is reply               & +0.13 \\
		Number of links        & +0.11 \\
		Number of times listed & +0.11 \\
		Number of statuses     & -0.20 \\
		\bottomrule
	\end{tabular}
	\caption{Social features that have the biggest impact on performance when
	removed from the model.}
	\label{tab:deletion-ablation}
\end{table}
\end{comment}

The number of followers a user has is often considered one of the measures of
her popularity.  While it is certainly not the only one or the ``best''
one~\cite{cha10measuring}, it is still fairly indicative of the user's
popularity/influence and much easier to collect than other ones (e.g., number
of mentions).  In the next experiment, we are interested in seeing how well our
system predicts what popular users (those with at least a certain number of
followers) will delete.  In addition, we look at how well our system works
for verified users (celebrities).  Arguably, predicting whether a celebrity or
a user with 10,000 followers will delete a tweet is a much more interesting
task than predicting if a user with 3 followers will do so.  To do this, we run
experiments where we only train and test on those users with the number of
followers in a certain range, or only on those users that are verified.  We can
see from Table~\ref{tab:deletion-groups} that the situation between groups is
very different.  While for users with less than 1,000 followers the performance
goes down, our system does much better on users that have lots of followers (it
is also interesting to note that the baseline is much higher for users with
more followers, which means that they are more likely to delete tweets in the
first place).  In fact, for users with more than 10,000 followers our system
achieves very good performance that it could actually be applied in a real
scenario.  For celebrities, results are somewhat lower, but still much higher
than for the whole training set.

\begin{table}
	\hspace*{-4mm}
	\centering
	\begin{tabular}{l r r r}
		\toprule
		User group & Here & Baseline & \# in test set \\
		\midrule
		Followers $< 1,000$               & 17.8 & 5.8   &  6.8M \\

		%Followers $\in [1k, 10k]$   & 42.0 & 6.6  \\ % & 640,098 \\
		%Followers $\in [10k, 100k]$ & 71.2 & 17.7 \\ % & 50,413  \\
		%Followers $> 100,000$       & 87.5 & 41.5 \\ % & 5,495   \\
		%Celebrities                 & 49.2 & 6.0  \\ % & 3,533   \\
		Followers $\in [1k, 10k]$   & 33.7 & 6.6  & 640k \\
		Followers $\in [10k, 100k]$ & 66.0 & 17.7 & 50k  \\
		Followers $> 100,000$       & 86.4 & 41.5 & 5.5k   \\
		Celebrities                 & 39.5 & 6.0  & 3.5k   \\
		% Lists $> 10$                    & 27.9 & & 1,429,782 \\
		\bottomrule
	\end{tabular}
	\caption{$F_1$ score for different groups of users.  The third column shows
	our results for named groups. The last column shows the
	number of users in the test set that fall into each category.}
	\label{tab:deletion-groups}
\end{table}

\section{Why are tweets deleted?}

One of the fundamental questions concerning deleted tweets is why are 
they deleted in the first place.  Is it the case that most of the deletion
notices that we see in the stream are there because users deleted their
accounts?  Or is it the case that most of the deleted tweets come from active
Twitter users who change their mind about posting a tweet (for one reason or
another)? Or are tweets deleted by Twitter because they are sent by spammers?
Here we try to answer these questions by looking at profiles of
users who deleted tweets.

We take the 200000 deleted tweets from the test set and query Twitter's API to
retrieve the account status of their author.  There are three possible
outcomes: the account still exists, the account exists but it is protected, or
the account does not exist any more.  Deleted tweets from the first type of
user are tweets that users manually delete and are probably the most
interesting case here.  Deleted tweets from users who have made their accounts
protected are probably not really deleted, but are only available to read for a
very small group of users.  The third case involves users who have had their
entire accounts deleted and thus none of their tweets are available any more.
While it is possible for a user to delete his account himself, it is much more
likely that these users are spammers and have had their accounts deleted by
Twitter.  Statistics about these three types of deletions are shown in
Table~\ref{tab:users-deleted}.  Most of the deleted tweets are genuine
deletions rather than a consequence of deleting spammers, showing that there is
much more to predicting deletions than simply predicting spam tweets.

\begin{table}[htb]
%	\hspace*{-4mm}
	\centering
	\begin{tabular}{l r r}
		\toprule
		Deletion type   & \% of tweets in test set & Accuracy \\
		\midrule
		Manual deletion & 85.2                        & 18.8     \\
		Protected       & 12.2                        & 17.5     \\
		Account deleted & 2.6                         & 29.5     \\
		\bottomrule
	\end{tabular}
	\caption{Proportion of different types of deletions and
	performance of our algorithm across these types.}
	\label{tab:users-deleted}
\end{table}

Given this classification of deletions, we are interested in finding out how our
approach performs across these different groups.  Is it the case that some
deletions are easier to predict than others?  In order to answer this question,
we test the performance of our system on the deleted tweets from these three
groups.  Because each of the three test sets now contains only positive
examples, we measure performance in terms of accuracy instead of $F_1$ score.
Note also that in this case accuracy is the same as recall.  The third column
of Table~\ref{tab:users-deleted} shows that i) predicting deletions that are a
result of deleted accounts (i.e., spotting spammers) is much easier than
predicting genuine deletions, and ii) predicting which tweets will become
protected is the hardest task.

Our manual analysis of the tweets discovered that a lot of deleted tweets
contained curse words, leading us to examine the relationship between cursing
and deletion in more detail.  Curse words are known to express negative
emotions~\cite{jay09utility}, which lead us to hypothesize that tweets which
contain curse words are more likely to be deleted.
%This was guided by the intuition that when people are angry they tend to not
%think before posting, and are thus likely to regret these tweets.  
In order to test this hypothesis, we calculate the probabilities of a tweet
being deleted conditioned on whether it contains a curse word.  We use a list
of 68 English curse words, and only consider English tweets from the test set.
We find that the probability of deletion given that the tweet contains a curse
word is 3.73\%, compared to 3.09\% for tweets that do not contain curse words.
We perform a two-sample z-test and find that the difference is statistically
significant at $p=0.0001$, which supports our hypothesis.

\section{Conclusion}

We have proposed a new task: predicting which messages on Twitter will be
deleted in the future.  We presented an analysis of the deleted messages on
Twitter, providing insight into the different reasons why people delete tweets.
To the best of our knowledge, we are the first to conduct such an analysis.
Our analysis showed, e.g., that tweets which contain swear words are more
likely to be deleted.  Finally, we presented a machine learning approach and
showed that for certain groups of users it can predict deleted messages with
very high accuracy.

%\section*{Acknowledgements}

\bibliographystyle{acl}
%\bibliography{bibtex.bib}

\begin{thebibliography}{}

\bibitem[\protect\citename{Bamman \bgroup et al.\egroup
  }2012]{bamman12censorship}
David Bamman, Brendan O'Connor, and Noah~A. Smith.
\newblock 2012.
\newblock {Censorship and Deletion Practices in Chinese Social Media}.
\newblock {\em First Monday}, 17(3).

\bibitem[\protect\citename{Bollen \bgroup et al.\egroup }2011]{bollen11twitter}
Johan Bollen, Huina Mao, and Xiaojun Zeng.
\newblock 2011.
\newblock Twitter mood predicts the stock market.
\newblock {\em Journal of Computational Science}, 2(1):1--8.

\bibitem[\protect\citename{Cha \bgroup et al.\egroup }2010]{cha10measuring}
Meeyoung Cha, Hamed Haddadi, Fabricio Benevenuto, and Krishna~P. Gummadi.
\newblock 2010.
\newblock Measuring user influence in {T}witter: The million follower fallacy.
\newblock In {\em 4th International AAAI Conference on Weblogs and Social Media
  (ICWSM)}, pages 10--17.

\bibitem[\protect\citename{Cortes and Vapnik}1995]{cortes95support}
Corina Cortes and Vladimir Vapnik.
\newblock 1995.
\newblock Support-vector networks.
\newblock {\em Machine learning}, 20(3):273--297.

\bibitem[\protect\citename{Crammer \bgroup et al.\egroup
  }2006]{crammer06online}
Koby Crammer, Ofer Dekel, Joseph Keshet, Shai Shalev-Shwartz, and Yoram Singer.
\newblock 2006.
\newblock {Online passive-aggressive algorithms}.
\newblock {\em The Journal of Machine Learning Research}, 7:551--585.

\bibitem[\protect\citename{Dabbish \bgroup et al.\egroup
  }2003]{dabbish03marked}
Laura Dabbish, Gina Venolia, and JJ~Cadiz.
\newblock 2003.
\newblock Marked for deletion: an analysis of email data.
\newblock In {\em CHI '03 extended abstracts on Human factors in computing
  systems}, CHI EA '03, pages 924--925, New York, NY, USA. ACM.

\bibitem[\protect\citename{Dabbish \bgroup et al.\egroup
  }2005]{dabbish05understanding}
Laura Dabbish, Robert~E. Kraut, Susan Fussell, and Sara Kiesler.
\newblock 2005.
\newblock Understanding email use: predicting action on a message.
\newblock In {\em Proceedings of the SIGCHI conference on Human factors in
  computing systems}, pages 691--700. ACM.

\bibitem[\protect\citename{Fan \bgroup et al.\egroup }2008]{fan08liblinear}
Rong-En Fan, Kai-Wei Chang, Cho-Jui Hsieh, Xiang-Rui Wang, and Chih-Jen Lin.
\newblock 2008.
\newblock {LIBLINEAR}: A library for large linear classification.
\newblock {\em Journal of Machine Learning Research}, 9:1871--1874.

\bibitem[\protect\citename{Jay}2009]{jay09utility}
Timothy Jay.
\newblock 2009.
\newblock The utility and ubiquity of taboo words.
\newblock {\em Perspectives on Psychological Science}, 4(2):153--161.

\bibitem[\protect\citename{O'Connor \bgroup et al.\egroup
  }2010]{oconnor10tweets}
Brendan O'Connor, Ramnath Balasubramanyan, Bryan~R. Routledge, and Noah~A.
  Smith.
\newblock 2010.
\newblock From tweets to polls: Linking text sentiment to public opinion time
  series.
\newblock In {\em Proceedings of the 4th International Conference on Weblogs
  and Social Media}, pages 122--129. The AAAI Press.

\bibitem[\protect\citename{Petrovi\'c \bgroup et al.\egroup
  }2011]{petrovic11rt}
{Sa\v sa} Petrovi\'c, Miles Osborne, and Victor Lavrenko.
\newblock 2011.
\newblock {RT to Win! Predicting Message Propagation in Twitter}.
\newblock In {\em Proceedings of ICWSM}.

\bibitem[\protect\citename{Ritterman \bgroup et al.\egroup
  }2009]{ritterman09using}
Joshua Ritterman, Miles Osborne, and Ewan Klein.
\newblock 2009.
\newblock Using prediction markets and twitter to predict a swine flu pandemic.
\newblock In {\em 1st International Workshop on Mining Social Media}.

\bibitem[\protect\citename{Tschang}2012]{tschang12analysis}
Chi-Chu Tschang.
\newblock 2012.
\newblock An analysis of sina weibo censorship using weiboscope search data.
\newblock http://partnews.mit.edu/.

\bibitem[\protect\citename{Tumasjan \bgroup et al.\egroup
  }2010]{tumasjan10predicting}
A.~Tumasjan, T.O. Sprenger, P.G. Sandner, and I.M. Welpe.
\newblock 2010.
\newblock {Predicting elections with Twitter: What 140 characters reveal about
  political sentiment}.
\newblock In {\em International AAAI Conference on Weblogs and Social Media,
  Washington, DC}.

\end{thebibliography}
\newcommand{\SortNoop}[1]{}

\end{document}